\documentclass[twocolumn,superscriptaddress,amsmath,amssymb,pra,longbibliography]{revtex4-1}

\ifx\pdfoutput\@undefined\usepackage[usenames,dvips]{color}
\else\usepackage[usenames,dvipsnames]{color}
\usepackage{graphicx}
\usepackage{bm}
\usepackage{amssymb}
\usepackage{hyperref}
\usepackage{color}

\usepackage{amsmath}

\usepackage[T1]{fontenc}
\usepackage[latin1]{inputenc}

\renewcommand{\vec}[1]{\mathbf{#1}}

\newcommand\Ev{\vec{E}}
\newcommand\Hv{\vec{H}}

\newcommand{\eps}{\varepsilon}
\newcommand{\epsh}{\hat{\eps}}
\newcommand{\muh}{\hat{\mu}}

\newcommand{\pcsadd}{Center for Theoretical Physics of Complex Systems, Institute for Basic Science, Daejeon 34126, Korea}

\usepackage{graphicx}
\usepackage{color}

\begin{document}

\title{Edge modes in 2D electromagnetic slab waveguides: analogues of acoustic plasmons}

\author{Daniel Leykam}
\affiliation{\pcsadd}
\affiliation{Basic Science Program, Korea University of Science and Technology, Daejeon 34113, Korea}
\author{Konstantin Y. Bliokh}
\affiliation{Theoretical Quantum Physics Laboratory, RIKEN Cluster for Pioneering Research, Wako-shi, Saitama 351-0198, Japan}
\author{Franco Nori}
\affiliation{Theoretical Quantum Physics Laboratory, RIKEN Cluster for Pioneering Research, Wako-shi, Saitama 351-0198, Japan}
\affiliation{Physics Department, University of Michigan, Ann Arbor, Michigan 48109-1040, USA}


\begin{abstract}
We analyze planar electromagnetic waves confined by a slab waveguide formed by two perfect electrical conductors. Remarkably, 2D Maxwell equations describing transverse electromagnetic modes in such waveguides are exactly mapped onto equations for acoustic waves in fluids or gases. We show that interfaces between two slab waveguides with opposite-sign permeabilities support 1D edge modes, analogous to surface acoustic plasmons at interfaces with opposite-sign mass densities. We analyze this novel type of edge modes for the cases of isotropic media and anisotropic media with tensor permeabilities (including hyperbolic media). We also take into account `non-Hermitian' edge modes with imaginary frequencies or/and propagation constants. Our theoretical predictions are feasible for optical and microwave experiments involving 2D metamaterials.
\end{abstract}

\maketitle

\section{Introduction}

Surface wave modes play crucial roles in metamaterials and topological wave systems \cite{Smith2004,Zayats2005,Alu2007,Bliokh2008,Hasan2010,Lu2014,Ozawa2019}. Such systems are intensively studied in both electromagnetism (optics) \cite{Smith2004, Alu2007, Lu2014, Ozawa2019, Wang2009, Rechtsman2013, Khanikaev2013, Slobozhanyuk2017} and acoustics 
\cite{Lu2009, Hussein2014, Ma2016, Cummer2016, Yang2015, Wang2015, Nash2015, He2016, Ni2019}. Surface electromagnetic waves between continuous isotropic media are also known as surface plasmon-polaritons, which appear at interfaces where the permeability or/and permittivity of the medium change their signs \cite{Zayats2005,Alu2007,Bliokh2008,Shadrivov2004,Kats2007,
Bliokh2019}. Similar modes can also appear at acoustic interfaces  \cite{Ambati2007,Park2011,Bliokh2019PRB}, but these are rather exotic, because the effective mass density must change its sign across the interface. Up to now there were very few studies of such acoustic surface plasmons. 

In this work we describe a novel type of surface electromagnetic mode, namely 1D edge modes at interfaces between two 2D slab waveguides (formed by perfect electric conductors) with different signs of permeability. Such modes have a three-fold interest. First, these can be generated in 2D metamaterials, which have a number of advantages as compared to bulk 3D metamaterials: more compact design, smaller losses, etc. Second, we show that Maxwell equations for waves in 2D slab waveguide are entirely analogous to the equations of acoustics, and the new electromagnetic modes we describe are analogues of acoustic plasmons at interfaces with different signs of the mass density \cite{Ambati2007,Park2011,Bliokh2019PRB}. Third, by considering anisotropic (uniaxial) permeability tensors we demonstrate the persistence of these edge modes in anisotropic media with elliptic or hyperbolic dispersion.

We first describe the general mapping between the equations of 2D electromagnetism and acoustics (Section \ref{sec2}). Then, we consider edge modes at interfaces between two slab waveguides with different parameters (Section \ref{sec3}). 
We consider both isotropic media 
and uniaxial anisotropic media, where permeabilities parallel and normal to the interface differ from each other. The latter case also includes hyperbolic media \cite{Menon2012,Poddubny2013}. Following the recent `non-Hermitian' approach to surface waves between continuous electromagnetic and acoustic media  \cite{Bliokh2019,Bliokh2019PRL}, 
we take into account edge modes with real and imaginary frequency and propagation constant. 

We argue that the system we describe is quite feasible for optical and microwave experiments, so that it can be used as a convenient platform for experimental studies (and possibly applications) of exotic surface/edge modes in wave systems.  

\begin{figure}
\includegraphics[width=0.9\columnwidth]{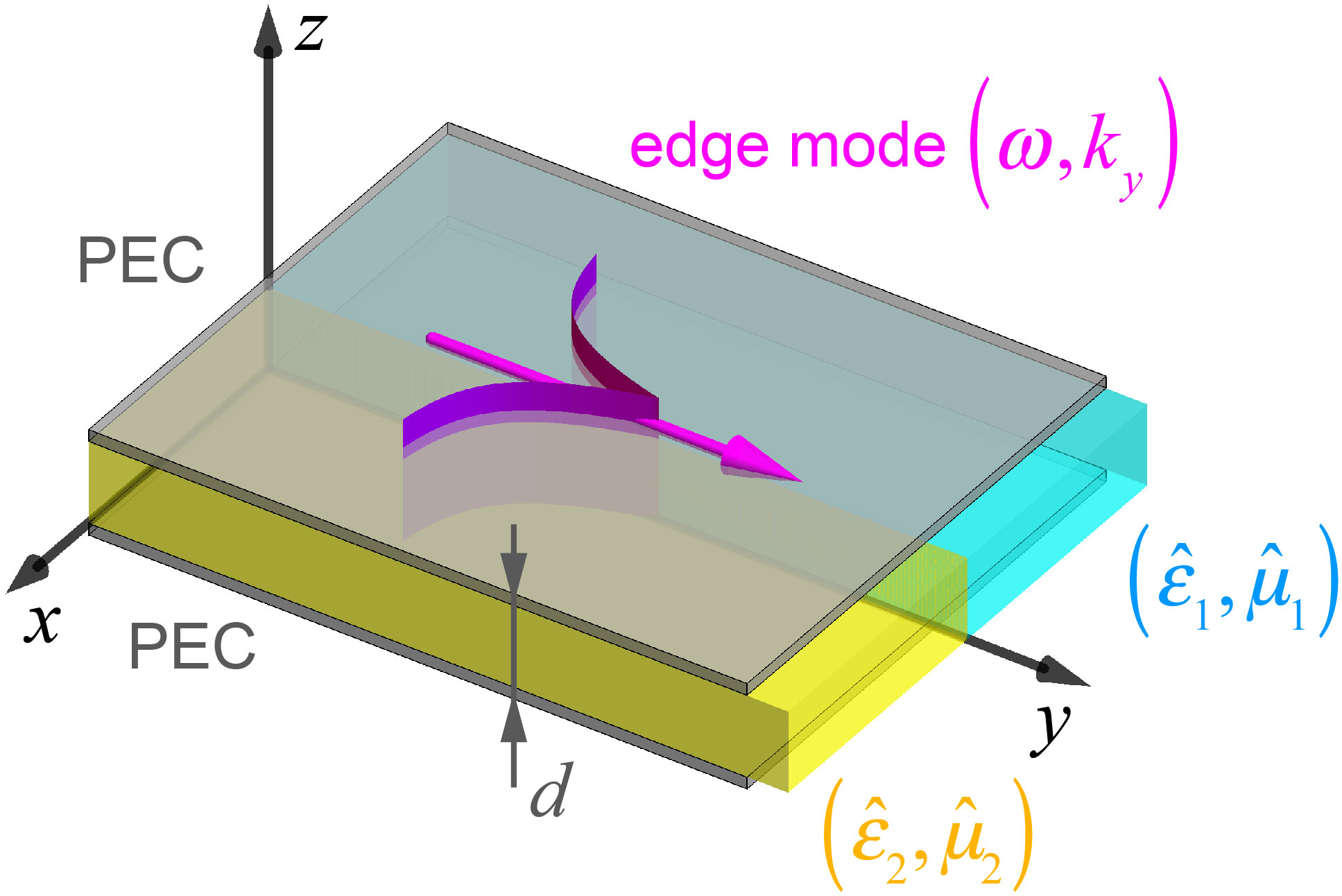}
\caption{System geometry: a thin slab waveguide embedded between two perfect electrical conductors (PECs). Inside the slab there is an interface between two different materials at $x=0$, which can support edge modes propagating in the $y$ direction. 
\label{Fig1}}
\end{figure}

\section{2D electromagnetism versus acoustics}
\label{sec2}

We consider  light propagation in a slab waveguide formed by two perfect electrical conductor (PEC) plates, shown in Fig.~\ref{Fig1}. We allow the medium between the plates to be anisotropic, described by lossless (i.e., Hermitian) permittivity and permeability tensors $(\epsh,\muh)$. Maxwell's equations for monochromatic light at frequency $\omega$ are:
\begin{align}
\label{eq1}
  \nabla \times \Ev = i\, \omega\, \hat{\mu}\, \Hv \,, \quad 
\nabla \times \Hv = -i\, \omega\, \hat{\eps}\, \Ev \,, \nonumber \\
 \nabla \cdot \left(\hat{\mu}\, \Hv\right) = \nabla \cdot \left(\hat{\eps}\, \Ev \right) = 0 \,.
\end{align}
We take PEC boundary conditions at the $z=0$ and $z=d$ planes, $(E_x,E_y,B_z)|_{z=0} = (E_x,E_y,B_z)|_{z=d} = 0$, and seek transverse electromagnetic (TEM) modes with field components independent of $z$, i.e., $(E_x,E_y,B_z) \equiv 0$. Then the two curl equations (\ref{eq1}) become 
\begin{align}
\label{eq2}
 \left(\partial_y E_z, -\partial_x E_z, 0 \right)^T = i\, \omega\, \muh \, \Hv \,, \nonumber\\
\left(\partial_y H_z, -\partial_x H_z, \partial_x H_y- \partial_y H_x \right)^T = -i\, \omega\, \epsh\, \Ev \,.
\end{align}
Note that for $\omega \ne 0$, the solutions of these two equations, if they exist, automatically satisfy the boundary condition $B_z = (\muh\, \Hv)_z = 0$ (from the first equation) and the divergence equations (\ref{eq1}): $\nabla \cdot \left(\muh\, \Hv \right) = \nabla \cdot \left( \epsh \, \Ev \right) = 0$ (by taking the divergence of the two equations and assuming $z$-independent fields).

We now assume that the anisotropy can only occur in the in-plane constitutive parameters, i.e.  
\[ \varepsilon_{zj} = \varepsilon_{jz} = \mu_{zj} = \mu_{jz} = 0\,,\]
where $j=x,y$. Then, valid solutions must have $H_z = 0$ (because $(\muh\, \Hv)_z = 0$), these are independent of $\mu_{zz}$, and 
only one component $\eps_{zz} \equiv \eps_z$ of $\epsh$ is relevant. 

Introducing the ``wavefunction'' $\Psi = \left(E_z,H_x,H_y\right)^T$, we can write the curl equations (\ref{eq2}) compactly as
\begin{equation} 
\label{eq3}
\left( \begin{array}{ccc} 0 & -\partial_y & \partial_x \\ -\partial_y & 0 & 0 \\ \partial_x & 0 & 0 \end{array} \right)\! \Psi = 
-i\,\omega\! \left( \begin{array}{ccc} \eps_z & 0 & 0 \\ 0& \mu_x & \mu_{xy} \\ 0 & \mu_{yx} & \mu_y \end{array} \right)\! \Psi \,,
\end{equation}
where $\mu_{xx} \equiv \mu_x$ and $\mu_{yy} \equiv \mu_y$. 

Remarkably, Eq.~(\ref{eq3}) is equivalent to acoustic equations for sound waves in fluids or gases \cite{LLFluid} in (2+1)D space-time. Indeed, introducing the ``pressure'' field $P = E_z$, ``velocity'' field ${\bf v} \equiv (v_x,v_y) = \bar{\bf z} \times {\bf H} = (-H_y,H_x)$ (where $\bar{\bf z}$ is the unit vector along the $z$-axis), as well as the ``compressibility'' $\beta=\eps_z$ and ``mass density'' $\hat{\rho} = \left(\begin{array}{cc} \mu_y &  -\mu_{yx} \\ -\mu_{xy} & \mu_x \end{array} \right)$ of the medium, we write Eq.~(\ref{eq3}) as follows (cf. \cite{Ambati2007,Bliokh2019PRB,Bliokh2019PRL}):
\begin{align}
\label{eq4}
\beta\, \partial_t P = - \nabla \cdot {\bf v} \,, \quad 
\hat{\rho}\, \partial_t {\bf v} = - \nabla P \,.  
\end{align}
Here we substituted $-i\,\omega \to \partial_t$ for monochromatic waves. 

Thus, one can use TEM slab waveguides to emulate wave propagation in 2D acoustic media with arbitrary parameters $\beta$ and $\hat{\rho}$. Note that filling the waveguide with 2D metamaterial structures, one can provide any desired parameters $\eps_z$ and $\mu_{jl}$, $(j,l)=\{x,y\}$, both positive and negative, at a given frequency $\omega$. 
This allows efficient electromagnetic emulation of acoustic metamaterials, including anisotropic ones with the tensor mass density $\hat{\rho}$.

Notably, the above 2D electromagnetic-to-acoustic mapping includes the main dynamical properties of the waves. In particular, the energy density $W$ and energy flux (Poynting vector) ${\bm\Pi}$ are consistent with both electromagnetic and acoustic theories \cite{Jackson,LLFluid,Bliokh2019PRB,Bliokh2019PRBII}:
\begin{align}
\label{eq5}
W = \frac{1}{4} \left( {\bf{E}}^* \epsh {\bf{E}} + {\bf{H}}^* \muh {\bf{H}} \right) 
= \frac{1}{4} \left(\beta |P|^2 + {\bf{v}}^* \hat{\rho} {\bf{v}} \right)\,, \nonumber \\
{\bm\Pi} = \frac{1}{2}\, {\rm Re}\! \left( {\bf{E}}^*\times {\bf H} \right)
= \frac{1}{2}\, {\rm Re}\! \left( P^* {\bf v} \right) \,.
\end{align}
Here we neglected, for the sake of simplicity, possible dispersion of the medium parameters. 

Furthermore, the quadratic forms, which describe the spin angular momentum density ${\bf S}$ in isotropic electromagnetic and acoustic media 
\cite{Shi2019, Bliokh2015, Bliokh2017PRL, Bliokh2019PRB, Bliokh2019PRBII,Burns2020}, are also equivalent:
\begin{align}
\label{eq6}
{\bf S} = \frac{1}{4\omega}\, {\rm Im}\! \left( {\bf{E}}^*\! \times \epsh {\bf{E}} + {\bf{H}}^*\! \times \muh {\bf{H}} \right) 
= \frac{1}{4\omega}\, {\rm Im}\! \left({\bf{v}}^*\! \times \hat{\rho} {\bf{v}} \right)\,. 
\end{align}
Note that in the system under consideration, the spin has purely magnetic origin (i.e., only the magnetic field can rotate), and it is purely transverse (i.e., orthogonal to the propagation $(x,y)$-plane): ${\bf S} \parallel \bar{\bf z}$ \cite{Bliokh2015,Bekshaev2015,Aiello2015,Neugebauer2018}.

\section{1D edge modes at isotropic and anisotropic interfaces}
\label{sec3}

We now consider an interface at $x=0$ between two slabs with different parameters $(\hat{\eps}_{1,2}, \hat{\mu}_{1,2}$) and seek solutions localized to the interface with localization lengths $\kappa_{1,2}>0$,
\begin{equation}
\Psi 
\propto \exp\!\left( i k_y y - |x| \kappa_{1,2} \right). 
\label{eq:edge_mode}
\end{equation}
References ~\cite{Ambati2007,Park2011,Bliokh2019PRB,Bliokh2019PRL} previously showed that interfaces between {\it isotropic} acoustic media where the sign of $\rho$ (i.e., $\mu$ is our system) changes exhibit surface modes analogous to electromagnetic surface plasmons, protected by a novel non-Hermitian bulk-boundary correspondence~\cite{Bliokh2019,Bliokh2019PRL}. While negative individual components of $\hat{\mu}$ can be readily implemented using microwave metamaterials such as split ring resonators, achieving isotropic negative $\hat{\mu}$ is much more challenging. Therefore we will consider the most general form of $\hat{\mu}$ to determine the conditions under which these edge modes persist in {\it anisotropic} media. 

Substituting Eq.~\eqref{eq:edge_mode} into Eq.~\eqref{eq3} and obtaining its eigenvectors yields the modal components 
\begin{equation}
\Psi = \left( \begin{array}{c} \omega(\mu_{xy,m}\mu_{yx,m} - \mu_{x,m} \mu_{y,m}) \\ 
-k_y \mu_{y,m} \mp i \mu_{xy,m} \kappa_{m} \\ 
k_y \mu_{yx,m} \pm i \mu_{x,m} \kappa_{m} \end{array}\right) e^{ i k_y y - \kappa_{m}|x|}\,. 
\label{eq:edge_mode2}
\end{equation}
Here $m=1,2$ denotes the medium, and the frequency satisfies
\begin{equation}
\omega^2 = \frac{-\mu_{x,m} \kappa_{m}^2 + \mu_{y,m} k_y^2 \pm i \kappa_{m} k_y\! \left(\mu_{xy,m} + \mu_{yx,m} \right)}{\eps_{z,m}\! \left(\mu_{x,m}\mu_{y,m} - \mu_{xy,m} \mu_{yx,m} \right)}, 
\label{eq:dispersion}
\end{equation}
where the upper sign ($+$) should be taken for medium 1 ($\kappa_1$), and lower sign ($-$) for medium 2 ($\kappa_2$).
%
Continuity of the tangential field components $E_z$ and $H_y$ at $x=0$ requires
\begin{equation}
\frac{\mu_{xy,1} \mu_{yx,1} - \mu_{x,1} \mu_{y,1}}
{\mu_{xy,2} \mu_{yx,2} - \mu_{x,2} \mu_{y,2}} 
= \frac{ k_y \mu_{yx,1} + i \mu_{x,1} \kappa_1}
{k_y \mu_{yx,2} - i \mu_{x,2} \kappa_2}\,. 
\label{eq:BCs}
\end{equation}
In the isotropic limit, where 
\[\mu_{xy,m}=\mu_{yx,m}=0 \quad {\rm and} \quad \mu_{x,m}=\mu_{y,m} \equiv \mu_m\,,\]
Eq.~\eqref{eq:BCs} reduces to the usual boundary condition for the TE surface mode $\tilde{\mu} \equiv \mu_2/\mu_1 = -\kappa_2/\kappa_1$~\cite{Shadrivov2004,Kats2007,Bliokh2019,Ambati2007,Park2011,Bliokh2019PRB}, which yields the topological bulk-boundary existence condition ${\rm sgn}\left(\tilde{\mu}\right) = -1$~\cite{Bliokh2019,Bliokh2019PRB}.

\begin{figure}
\includegraphics[width=0.68\columnwidth]{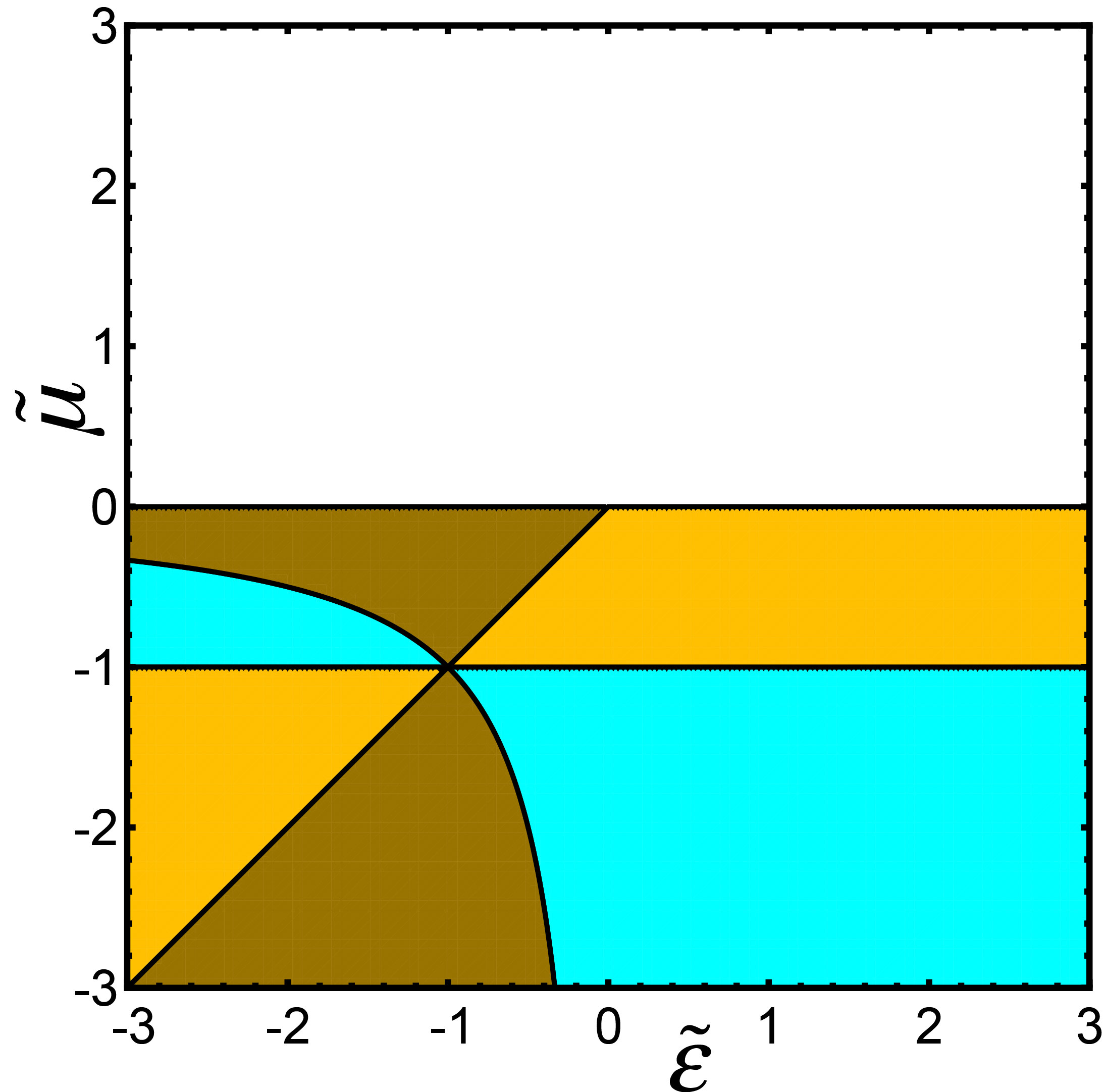}
\caption{Phase diagram of real-frequency edge modes supported by interfaces between a vacuum (blue; $\varepsilon_{z,1}=\mu_{x,1}=\mu_{y,1}=1$) or metal (yellow/brown; $\varepsilon_{z,1}=-1, \mu_{x,1}=\mu_{y,1}=1$) and an isotropic metamaterial with $\mu_{x,2} = \mu_{y,2} = \tilde\mu$ (cf. Refs.~\cite{Bliokh2019,Bliokh2019PRL}). Yellow indicates propagating (real $k_y$) solutions, while brown indicates evanescent (imaginary $k_y$) solutions.}
\label{fig2}
\end{figure}

In the general case of anisotropic media it does not seem possible to formulate a topological bulk-boundary correspondence because the condition for localized surface waves, $\kappa_1/\kappa_2 > 0$, cannot be expressed simply in terms of the properties of the interface, but rather depends implicitly on the bulk dispersion relations of the two media via the appearance of $k_y$ on the right hand side of Eq.~\eqref{eq:BCs}.
The problematic $k_y$ terms can be eliminated if the principal axes of the media are aligned with our coordinate axes, such that $\hat{\mu}$ becomes diagonal: 
\[ \mu_{xy,m} = \mu_{yx,m} = 0\,. \] 
In this case the boundary condition (\ref{eq:BCs}) reduces to
\begin{equation}
\tilde{\mu}_y \equiv \frac{\mu_{y,2}}{\mu_{y,1}} \ = -\frac{\kappa_2}{\kappa_1}\,. 
\label{eq:BC2}
\end{equation} 
This yields the existence condition in the topological form 
${\rm sgn}\left(\tilde{\mu}_y\right) = -1$,
i.e., localized edge modes can exist provided the sign of $\mu_y$ changes across the interface. 

The explicit solutions for the surface wave propagation constant and frequency obtained using Eqs.~(\ref{eq:edge_mode2}), (\ref{eq:dispersion}), and (\ref{eq:BC2}) are:
\begin{align}
&\frac{\mu_{y,1}}{\mu_{x,1}}\, k_y^2 =    \frac{\tilde\mu_{x}(\tilde\varepsilon_{z} - \tilde\mu_{y})}
{\tilde\varepsilon_{z}\tilde\mu_{x} - 1}\,
\kappa_1^2\,, \label{eq:kysurf} \\ 
&\varepsilon_{z,1} \mu_{y,1}\, \omega^2 =  
\frac{1 - \tilde\mu_{x} \tilde\mu_{y} }
{\tilde\varepsilon_{z} \tilde\mu_{x} - 1}\,
\kappa_1^2\,,
\label{eq:omsurf} 
\end{align}
where $\tilde\eps_z \equiv \eps_{z,2}/\eps_{z,1}$ and $\tilde\mu_{x} \equiv \mu_{x,2}/\mu_{x,1}$. Expressions (\ref{eq:BC2})--(\ref{eq:omsurf}) generalize the isotropic TE surface plasmon solutions discussed in Refs.~\cite{Shadrivov2004,Kats2007,Bliokh2019} and their acoustic analogues \cite{Ambati2007,Park2011,Bliokh2019PRB,Bliokh2019PRL}. 
Akin to the isotropic case, the edge modes can have either real or imaginary frequency $\omega$ or/and propagation constant $k_y$ \cite{Bliokh2019,Bliokh2019PRL}. In what follows, we will focus on the modes with real $\omega$ and either real or imaginary $k_y$. Such modes have physical sense of propagating and evanescent surface waves and can be observed experimentally.
For example, Fig.~\ref{fig2} shows the phase diagram for the existence of propagating and evanescent real-frequency edge modes for the simplest case of an interface between a vacuum ($\varepsilon_{z,1} = \mu_{x,1} = \mu_{y,1} = 1$) or a metal ($\varepsilon_{z,1} = -1, \mu_{x,1} = \mu_{y,1} = 1$) and an {\it isotropic} slab metamaterial with $\mu_{x,2} = \mu_{y,2} = \tilde\mu$, which reproduces the previously-demonstrated phase diagrams for 3D Maxwell's equations and acoustic waves~\cite{Bliokh2019,Bliokh2019PRL}. 

\begin{figure}
\includegraphics[width=0.68\columnwidth]{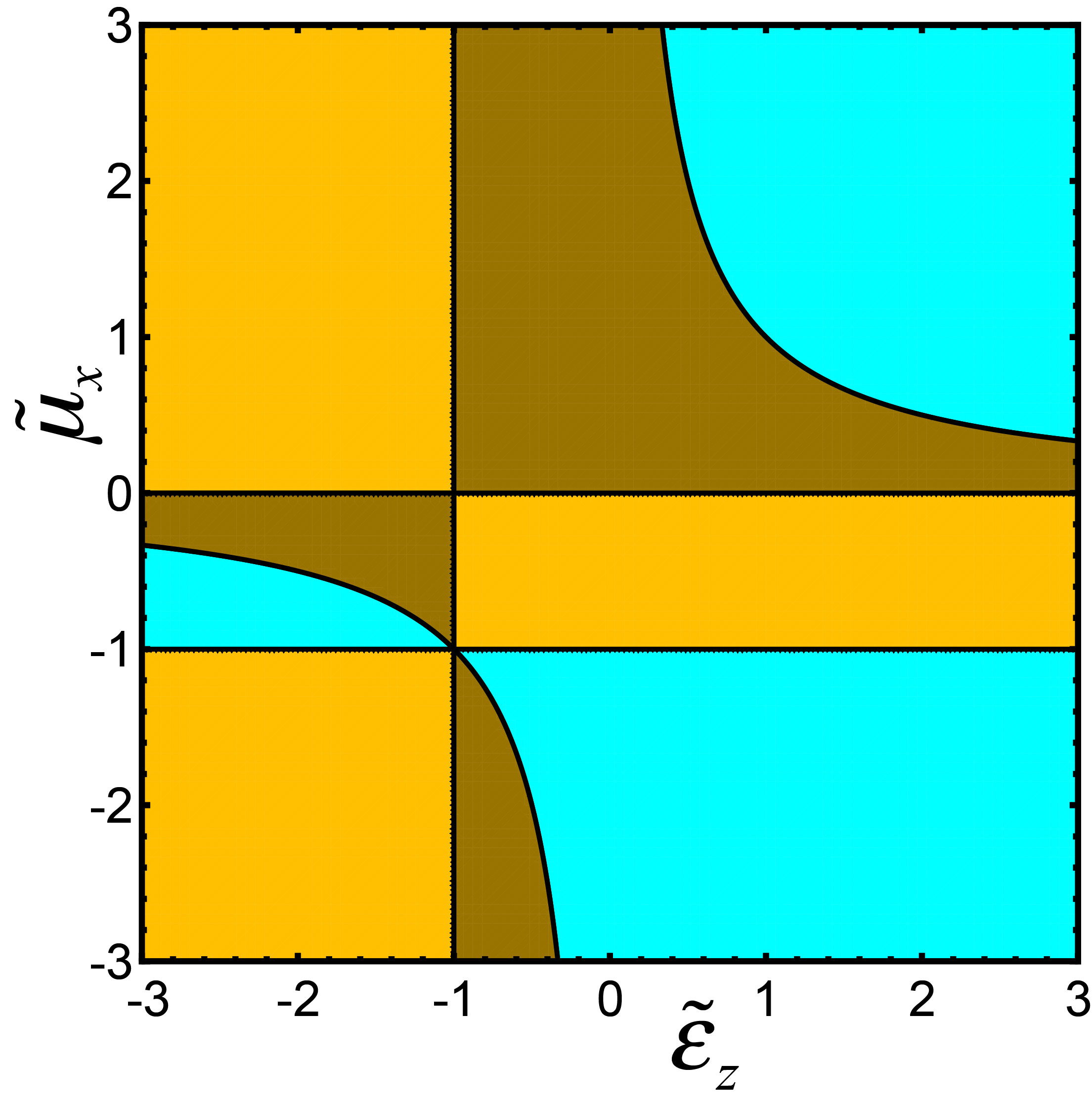}
\caption{Phase diagram of real-frequency edge waves supported by interfaces between a vacuum (blue; $\varepsilon_{z,1}=\mu_{x,1}=\mu_{y,1}=1$) or metal (yellow/brown; $\varepsilon_{z,1}=-1, \mu_{x,1}=\mu_{y,1}=1$) and an anisotropic metamaterial with $\mu_{y,2} = -1$. Yellow indicates propagating (real $k_y$) solutions, while brown indicates evanescent (imaginary $k_y$) solutions. The sign of $\tilde\mu_x$ determines whether the medium 2 has elliptic ($\tilde\mu_x <0$) or hyperbolic ($\tilde\mu_x > 0$) dispersion.}
\label{fig3}
\end{figure}

\begin{figure}
\includegraphics[width=0.75\columnwidth]{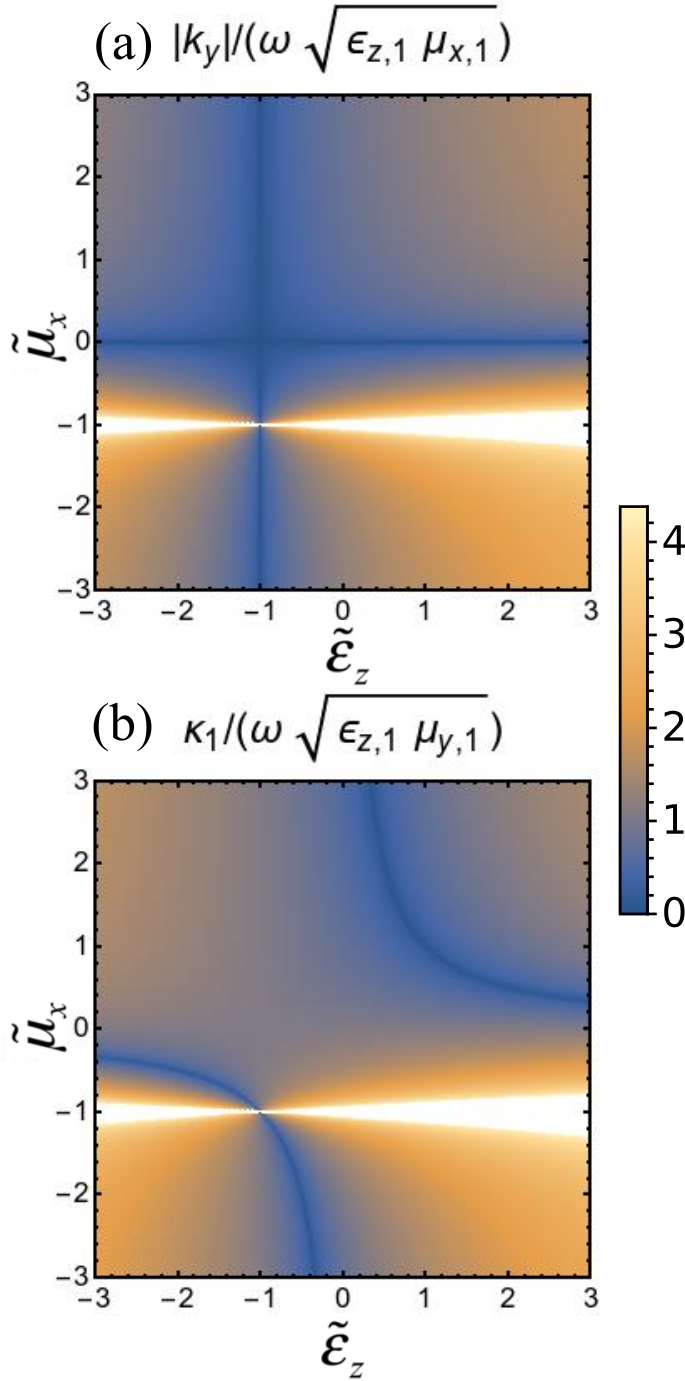}
\caption{Normalized edge mode propagation constant $k_y$ (a) and localization length $\kappa_1$ (b) for the modes shown in Fig.~\ref{fig3}, obtained using Eqs.~(\ref{eq:kysurf},\ref{eq:omsurf}). Boundaries between different types of solutions (e.g. propagating and evanescent) correspond to zeros or divergences of $k_y$, $\kappa_{1}$, or $\kappa_2 = -\tilde{\mu}_y \kappa_1$.}
\label{figR1}
\end{figure}

Figure~\ref{fig3} shows the phase diagram for the edge modes for the case when medium 2 is an {\it anisotropic metamaterial} with $\mu_{y,2} = -1$. Thus, the existence condition (\ref{eq:BC2}) is satisfied, while the sign of $\mu_{x,2}$ determines whether the medium 2 has {\it elliptic} ($\tilde\mu_{x}<0$) or {\it hyperbolic} ($\tilde\mu_{x}>0$) dispersion \cite{Menon2012,Poddubny2013}. We observe that propagating edge modes can persist for arbitrary anisotropy in the permeability tensor $\hat{\mu}$, provided the permittivity $\tilde\eps_z$ is chosen appropriately. Furthermore, interfaces between metal and hyperbolic metamaterials can also support evanescent edge modes \cite{Bliokh2019,Bliokh2019PRL}, which have never been observed so far.

According to Eqs.~(\ref{eq:kysurf},\ref{eq:omsurf}), transitions between the different types of surface modes can occur in three different ways: (i) Via a simultaneous divergence of $k_y$ and $\kappa_{1,2}$, occurring when $\tilde{\mu}_x \tilde{\mu}_y = 1$. Spatially-nonlocal corrections to the material dispersion will become important near this limit~\cite{PhysRevLett.124.153901}. (ii)  $k_y$ vanishes while $\kappa_{1,2}$ remain finite, occurring when $\tilde{\varepsilon}_z = \tilde{\mu}_y$ or $\tilde{\mu}_x = 0$ and corresponding to the edge mode's phase velocity vanishing. (iii) $\kappa_{1,2}$ vanishes while $k_y$ remains finite, occurring when $\tilde{\varepsilon}_z \tilde{\mu}_x = 1$ and corresponding to the edge mode delocalizing. Fig.~\ref{figR1} shows that each of scenarios can be observed in the $(\tilde{\varepsilon}_z,\tilde{\mu}_x)$ phase diagram of Fig.~\ref{fig3}.

From Eqs.~(\ref{eq:BC2})--(\ref{eq:omsurf}) and Figs.~\ref{fig2}--\ref{figR1}, we summarize the main differences of the anisotropic as compared to isotropic cases: 

(i) The topological condition for the existence of the edge mode in the isotropic case, ${\rm sgn} (\tilde\mu) = -1$ \cite{Bliokh2019,Bliokh2019PRL}, involves the longitudinal permeability in the anisotropic case: ${\rm sgn} (\tilde\mu_y) = -1$. 

(ii) The sign of the refractive index of medium 1, $n_1 =\sqrt{\eps_1\mu_1}$, which determines whether the edge mode has real or imaginary frequency $\omega$ in the isotropic case \cite{Bliokh2019,Bliokh2019PRL}, is replaced by the longitudinal refractive index $n_{y,1} = \sqrt{\eps_{z,1} \mu_{y,1}}$.

(iii) The transition from elliptic to hyperbolic dispersion in the anisotropic medium, controlled by $\tilde\mu_x$, swaps the propagating (real $k_y$) and evanescent (imaginary $k_y$) edge modes. 

Moreover, we emphasize that Eqs.~(\ref{eq:BC2})--(\ref{eq:omsurf}) depend only on the relative material parameters (up to a rescaling of $k_y$ and $\omega$), hence interfaces between two anisotropic metamaterials will exhibit qualitatively similar features.

\begin{figure}
\includegraphics[width=0.68\columnwidth]{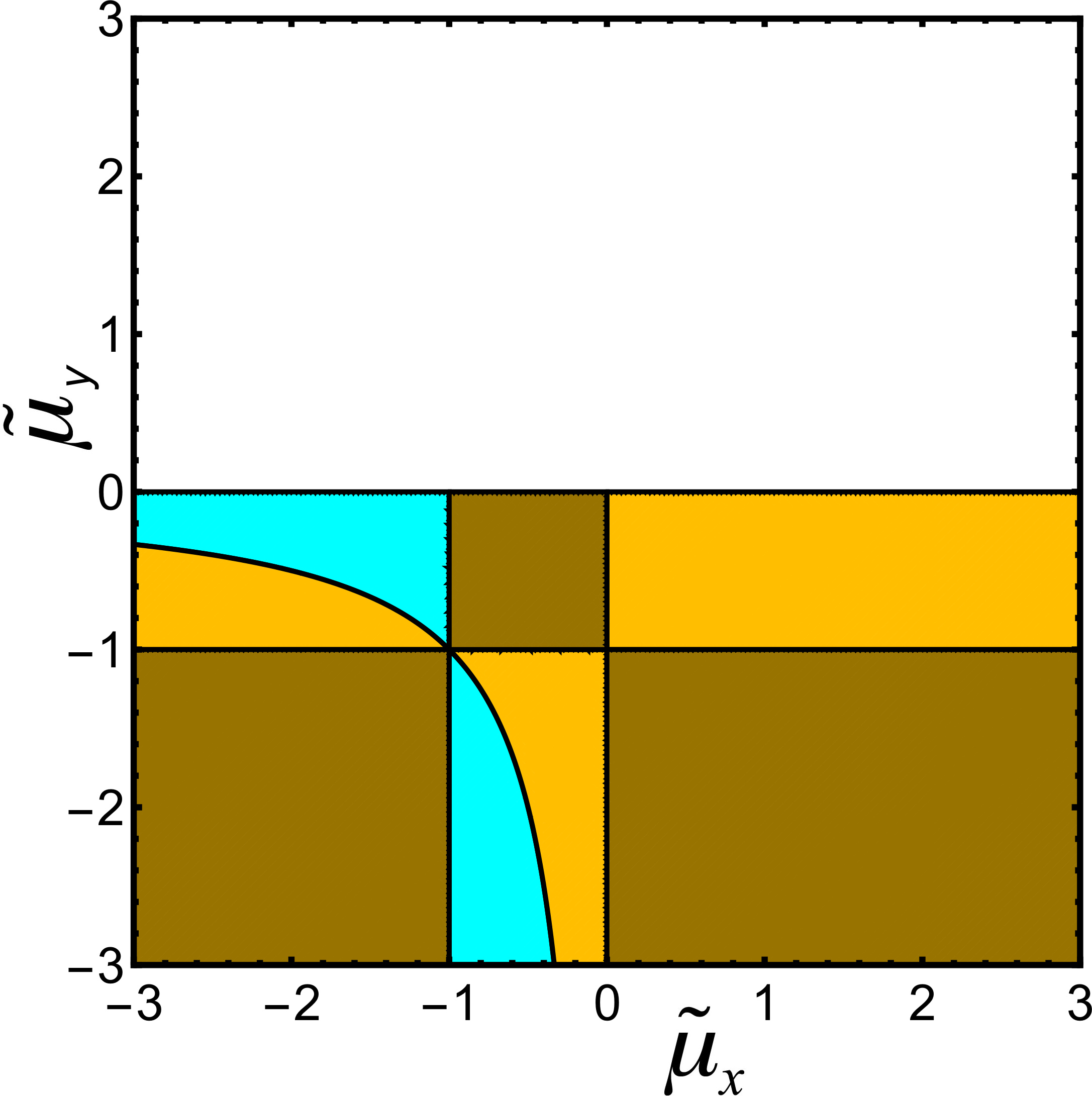}
\caption{Phase diagram of real-frequency edge waves supported by interfaces between a vacuum (blue; $\varepsilon_{z,1}=\mu_{x,1}=\mu_{y,1}=1$) or metal (yellow/brown; $\varepsilon_{z,1}=-1, \mu_{x,1}=\mu_{y,1}=1$) and an anisotropic metamaterial with $\eps_{z,2} = -\eps_{z,1}$. Yellow indicates propagating (real $k_y$) solutions, while brown indicates evanescent (imaginary $k_y$) solutions.}
\label{fig4}
\end{figure}

For completeness, Fig.~\ref{fig4} shows the phase diagram for edge modes as a function of $(\tilde{\mu}_x,\tilde{\mu}_y)$, for fixed $\tilde{\eps}_z = -1$. For this value of $\tilde{\eps}_z$ the vacuum interface supports real frequency edge modes over a narrower parameter region compared to a metallic interface. The latter also exhibits transitions between propagating and evanescent edge modes when $\tilde{\mu}_y = \tilde{\eps}_z$, which does not coincide with any change in the topology of the isofrequency surfaces of medium 2.

\section{Conclusions}

We have studied monochromatic electromagnetic waves in slab waveguides formed by two perfect electrical conductors, showing that Maxwell's equations governing transverse electromagnetic modes map exactly onto the two-dimensional acoustic equations for sound waves in fluids or gases. We derived conditions under which boundaries between two slab waveguides with different permittivities and anisotropic permeabilities support edge modes (similar to surface plasmons) protected by a topological bulk-edge correspondence. 

The system considered in our work provides a simple platform to emulate `acoustic surface plasmons' at interfaces where the mass density changes its sign \cite{Ambati2007,Park2011,Bliokh2019PRB,Bliokh2019PRL}, based on the analogy between the acoustic compressibility and mass density ($\beta, \hat{\rho}$) and the electromagnetic permittivity and permeability $(\varepsilon,\hat{\mu})$.
Moreover, this system can serve as an efficient platform for the experimental realization of recently-proposed non-Hermitian (evanescent) surface modes with imaginary propagation constants \cite{Bliokh2019,Bliokh2019PRB}. Indeed, we have shown that it does not require isotropic 3D metamaterials with negative permittivity or permeability, and it is sufficient to provide the negative sign of only one component of the permeability tensor, which can be readily implemented using hyperbolic metamaterials. For example, anisotropic negative permeability can be achieved at the polariton resonance of certain magnetic materials, or using metamaterials such as split ring resonators at microwave frequencies or multilayer fishnet structures in the near infrared~\cite{Poddubny2013}. Analogously, acoustic wave systems with a negative component of the mass density $\hat{\rho}$ have been demonstrated using arrays of thin plates~\cite{Ma2016,PhysRevLett.115.254301}.

We focused on anisotropic metamaterials described by a real permeability tensor $\hat{\mu}$. Another interesting class of anisotropic media are gyrotropic materials described by $\mu_x = \mu_y = \mu$ and $\mu_{xy} = -\mu_{yx} = i \alpha \mu$, which can be generated by placing a ferrite material between the plates and applying a magnetic field parallel to the $z$ axis. A similar configuration was used in the experiment of Ref.~\cite{Wang2009}, which, however, focused on a photonic crystal band structure rather than the low frequency response described by an homogeneous effective medium. References~\cite{Davoyan2013,Silveirinha2015,Horsley2018,UnidirectionalMaxwellianspinwaves} have described the unidirectional surface waves at boundaries between different classes of gyrotropic media. It will be interesting to study whether the bulk-boundary correspondence introduced in Ref.~\cite{Bliokh2019,Bliokh2019PRL} can be extended to these gyrotropic surface waves.

The edge modes we have considered may form the basis for novel subwavelength waveguides and resonators. Particularly interesting are the edge modes supported by interfaces between regular metals with $\varepsilon<0$ and negative index metals with $\mu < 0$, which can support propagating edge modes despite both bulk materials being metallic. Thereby losses due to bending of the waveguide may be completely eliminated. By modulating the material parameters parallel to the interface to alternate between propagating (real $k_y$) and evanescent (imaginary $k_y$) edge modes, one may also create highly localized resonant modes. Unavoidable metamaterial losses may be reduced by considering heterostructure waveguides formed by a thin film of one medium embedded within another~\cite{Tsakmakidis2006,Ishikawa2009}. Whether the modes of such thin film waveguides can be related to topological properties of the bulk media is another interesting question for future research.

\begin{acknowledgments}
We thank Alexey Slobozhanyuk and Pavel Belov for fruitful discussions. This work was partially supported by 
NTT Research, Army Research Office (ARO) (Grant No. W911NF-18-1-0358), Japan Science and Technology Agency (JST) (via the CREST Grant No. JPMJCR1676), Japan Society for the Promotion of Science (JSPS) (via the KAKENHI Grant No. JP20H00134, and the grant JSPS-RFBR Grant No. JPJSBP120194828), the Grant No. FQXi-IAF19-06 from the Foundational Questions Institute Fund (FQXi), a donor advised fund of the Silicon Valley Community Foundation, and the Institute for Basic Science in Korea (IBS-R024-Y1).
\end{acknowledgments}

\bibliography{bibA}

\end{document}